\pdfoutput=1
\documentclass[letterpaper, 10 pt, conference]{ieeeconf}  % Comment this line out if you need a4paper

\IEEEoverridecommandlockouts                              % This command is only needed if 
\usepackage{mathrsfs}
\usepackage[font={small}]{caption}
\usepackage{scalerel}
\usepackage{url}
\usepackage{bbold}
\usepackage{amsfonts}
\usepackage{amsmath,amssymb,amsfonts}
\usepackage{graphicx}
\usepackage{wrapfig}
\usepackage{mathrsfs} 
\usepackage{algorithm,algorithmic}
\usepackage{array}
\usepackage{times}
\usepackage{url}
\usepackage{subfigure}
\usepackage{cite}
\usepackage{upgreek}
\usepackage{float}
\usepackage{longtable}
\usepackage{color}
\usepackage{wasysym}
\usepackage{grffile}
\usepackage{stackengine}
\newcommand\xrowht[2][0]{\addstackgap[.5\dimexpr#2\relax]{\vphantom{#1}}}

\allowdisplaybreaks

\newcommand\numberthis{\addtocounter{equation}{1}\tag{\theequation}}

\newcommand{\0}{\mathbb{0}}

\newcommand{\E}{\mathbb{E}}
\newcommand{\R}{\mathbb{R}}

\newcommand{\xb}{\mathbf{x}}
\newcommand{\yb}{\mathbf{y}}

\newcommand{\Bb}{\mathbf{B}}

\newcommand{\Ib}{\mathbf{I}}

\newcommand{\Qb}{\mathbf{Q}}

\newcommand{\Xb}{\mathbf{X}}

\newcommand{\Zb}{\mathbf{Z}}

\newcommand{\Phib}{\boldsymbol{\Phi}}

\newcommand{\omegab}{\boldsymbol{\omega}}

\newcommand{\nub}{\boldsymbol{\nu}}
\newcommand{\thetab}{\boldsymbol{\theta}}

\newcommand{\Dc}{\mathcal{D}}

\newcommand{\Nc}{\mathcal{N}}

\newcommand{\Uc}{\mathcal{U}}

\newcommand{\qh}{\widehat{q}}
\newcommand{\Rh}{\widehat{R}}

\newcommand{\qbh}{\widehat{\mathbf{q}}}

\newcommand{\norm}[1]{\left\lVert#1\right\rVert}
\newcommand{\tr}[1]{\text{Tr}\left[#1\right]}
\newcommand{\inn}[1]{\left<#1\right>}

\newcommand{\ind}[1]{\mathbf{1}\left(#1\right)}
\newcommand{\ex}[1]{\E\left[#1\right]}

\newtheorem{problem}{Problem}

\newtheorem{remark}{Remark}

\title{\LARGE \bf Cell Association via Boundary Detection: \\ A Scalable Approach Based on Data-Driven Random Features}

\author{Yinsong Wang $^1$, Hessam Mahdavifar $^2$,  Kamran Entesari $^1$,  and Shahin Shahrampour $^1$ 
\thanks{$^1$Y. Wang, K. Entesari, and S. Shahrampour are with Texas A\&M University, College Station, TX 77843, USA. 
        {\tt\small email:\{gritti,kentesar,shahin\}@tamu.edu}.}%
\thanks{$^2$H. Mahdavifar is with the University of Michigan, Ann Arbor, MI 48109, USA. 
        {\tt\small email:hessam@umich.edu}.}%
}

\begin{document}

\maketitle
\thispagestyle{empty}
\pagestyle{empty}

%%%%%%%%%%%%%%%%%%%%%%%%%%%%%%%%%%%%%%%%%%%%%%%%%%%%%%%%%%%%%%%%%%%%%%%%%%%%%%%%
\begin{abstract}
The problem of cell association is considered for cellular users present in the field. This has become a challenging problem with the deployment of 5G networks which will share the sub-6 GHz bands with the legacy 4G networks. Instead of taking a network-controlled approach, which may not be scalable with the number of users and may introduce extra delays into the system, we propose a scalable solution in the physical layer by utilizing data that can be collected by a large number of spectrum sensors deployed in the field. More specifically, we model the cell association problem as a nonlinear boundary detection problem and focus on solving this problem using randomized shallow networks for determining the boundaries for location of users associated to each cell. We exploit the power of data-driven modeling to reduce the computational cost of training in the proposed solution for the cell association problem. This is equivalent to choosing the right basis functions in the shallow architecture such that the detection is done with minimal error. Our experiments demonstrate the superiority of this method compared to its data-independent counterparts as well as its computational advantage over kernel methods. 
\end{abstract}

%%%%%%%%%%%%%%%%%%%%%%%%%%%%%%%%%%%%%%%%%%%%%%%%%%%%%%%%%%%%%%%%%%%%%%%%%%%%%%%%
\section{Introduction}

Wireless networks are rapidly growing in size, are becoming increasingly distributed, and are granted access to increasingly wider frequency spectrum. In the next generation of wireless cellular networks, namely 5G,
tens of small cells, hundreds of mobile users demanding ultra-high data rates, and thousands of Internet-of-Things (IoT) and machine-type communication (MTC) devices will be all operating within the coverage of one single cell \cite{3gpp-5g-2}. Furthermore, 5G systems will be deployed across the extensive available frequency
spectrum including the bands below 6\,GHz, the radio frequency
(RF) band, as well as bands around 30\,GHz, the millimeter wave (mm-wave) band \cite{3gpp-5g-3}. However, one of the grand challenges in deploying 5G systems is to ensure the coexistence with current 4G systems in the foreseeable future. Tight interworking between 4G and 5G and dynamic spectrum sharing between these two systems
are key to the smooth migration towards 5G \cite{lin2019debunking}. This necessitates decentralized and real-time data-driven techniques for user management, including cell association, in cellular networks that can be properly scaled with the massive number of users in future systems. 
 
 In general, in heterogeneous networks, including coexisting 4G/5G networks in sub-6 GHz bands, classical methods for cell association based only on received signal power are not interference aware and hence, may lead to huge traffic load imbalance \cite{hossain2014evolution}. To this end, various solutions have been proposed in the litrature in order to address interference awareness \cite{sangiamwong2011investigation}, traffic load awareness \cite{guvenc2011capacity}, and resource awareness \cite{oh2012cell}. However, such network-controlled centralized algorithms for user management, which would need to be coordinated between 4G and 5G networks as well, introduce extra delay to the system and may not guarantee quality of service (QoS) constraints especially for the cell-edge users. Instead of taking such a network-controlled approach, we model the cell association problem as a boundary detection problem and propose a scalable data-driven solution in the physical layer for this problem by utilizing data that can be collected by a large number of low-cost spectrum sensors deployed in the field. Deploying such spectrum sensors has been considered in various scenarios such as cognitive radio networks \cite{cabric2004implementation,ghasemi2008spectrum,sepidband2015cmos}. 
 
 The problem of coverage detection has been modeled and studied as a boundary detection problem in various contexts including sensor networks \cite{wang2006boundary} and cognitive radio networks \cite{yang2010cooperative,ding2013kernel}. However, in order to make this model applicable to cellular networks, and in particular to coexisting 4G/5G networks, in this paper, we extend the model in several important respects. We consider two base stations (BS) which can have different powers and coverage. Also, the declaration by each of the sensors in the field can have three different possibilities, i.e., whether there is a coverage in which case the stronger BS is declared or there is no sufficiently strong coverage from either of BSs. The analysis can be extended to cases with more than two BSs and, consequently, more than three possibilities for the declared labels in a straightforward way. 
 
 Of particular relevance to our paper is the work of \cite{ding2013kernel}, where the boundary detection problem has been tackled using kernel-based methods. While kernel methods are popular modeling tools in machine learning and statistics \cite{friedman2001elements}, they suffer from scalability issues, i.e., they scale poorly with respect to the number of sensors. We will use randomized features \cite{rahimi2007random,rahimi2009weighted} to reduce the computational cost of kernel methods. In particular, we adapt the data-driven random feature method of \cite{wang2019general} for the considered boundary detection problem. We show in numerical experiments that the data-driven solution consistently outperforms the data-independent methods (in prediction accuracy) and enjoys a much lower training cost compared to kernel methods.

\begin{figure*}[t!]
\centering
 \includegraphics[width=0.85\textwidth]{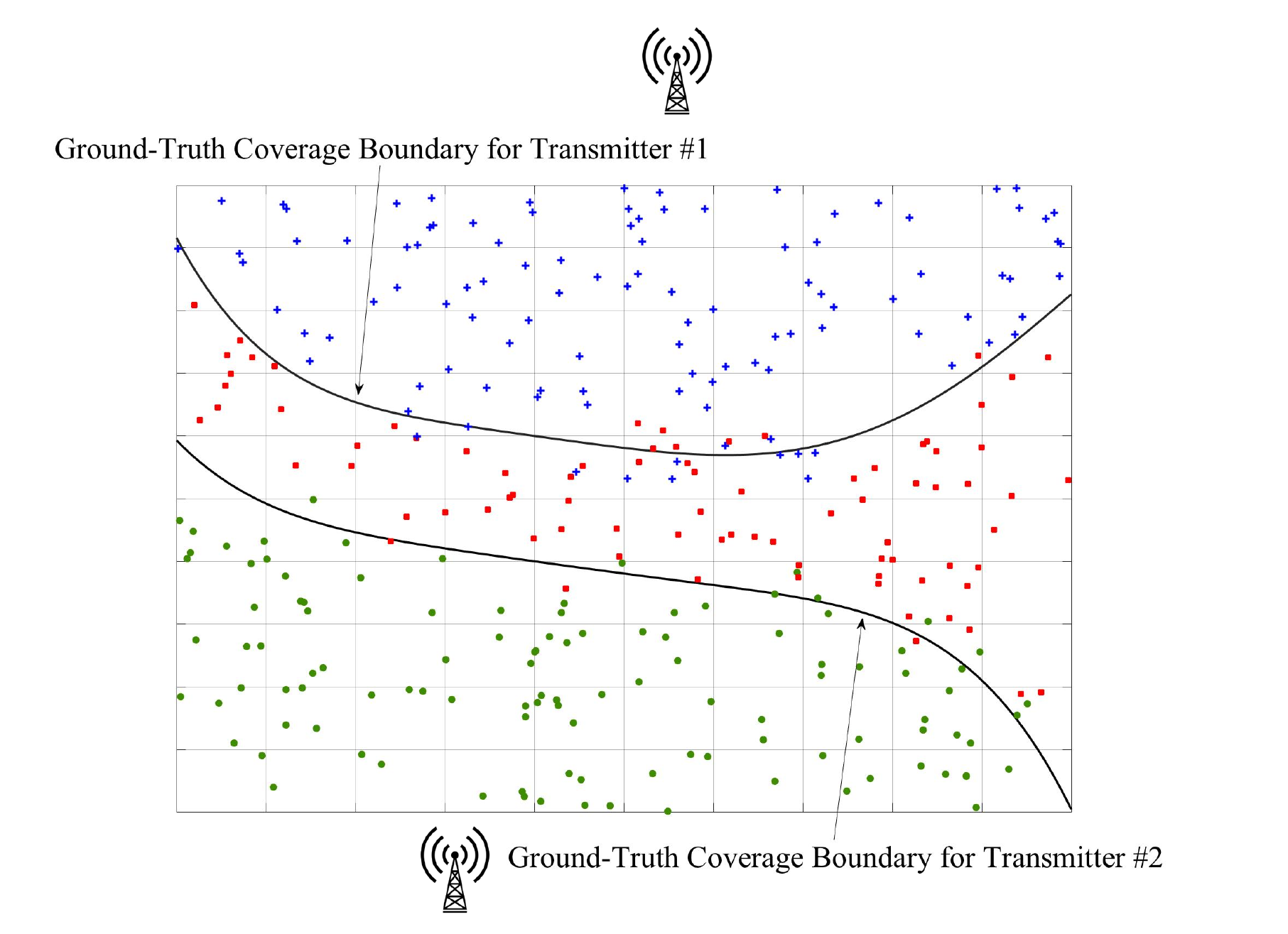}
\caption{\it The boundary detection problem with two transmitters (base stations).}
\label{transmitters}
\end{figure*}

\section{Problem Statement}
In this section, we describe the boundary detection problem for the case of two base stations (BSs). Note that we refer to the cellular transmitter, which is often called eNB in 4G networks and gNB in 5G networks, as the BS regardless of whether it belongs to 4G or 5G network. The described idea here is generalizable to more than two BSs and the simplification is only for the presentation clarity. Let us now consider Fig. \ref{transmitters} as an example of the boundary detection problem with two BSs (namely, BS1 and BS2). Each BS has a ground-truth coverage boundary between its corresponding \textit{covered} area and the rest of the plane. As illustrated in Fig.\,\ref{transmitters}, the ground-truth boundaries may be highly nonlinear (irregular). Irregular radio \textit{coverage} model (for the case of one BS) was proposed by \cite{ding2013kernel} to generalize the setup of \cite{yang2010cooperative}. This was motivated by the fact that signal attenuation due to obstructions (e.g., buildings) can affect the shape of the boundaries. In the considered cases with more than one BS, the interference from the neighboring cells, especially in areas in between also referred to as cell edge, depends on the power and the traffic in these cells. Consequently, this becomes another factor contributing to the non-linearity of the \textit{coverage} boundaries.

Consider now $n$ spectrum sensors, randomly distributed in a 2D area. Sensor $i\in \{1,\ldots,n\}$ is located at $\xb_{i}=[\xb_i(1),\xb_i(2)]^\top \in \R^2$. Each sensor is equipped with an energy detector, identifying the \textit{coverage} from transmitters. In this example, this identification at a particular sensor can lead to three possibilities: (i) the sensor senses \textit{strong} coverage from BS1 (blue/plus points); (ii) the sensor senses \textit{strong} coverage from BS2 (green/circle points); (iii) the sensor does not declare a sufficiently \textit{strong} coverage from either BSs (red/square points). As illustrated in Fig. \ref{transmitters}, some of the declarations are prone to errors due to the hardware constraints of the spectrum sensors and radio channel randomness. Indeed, no knowledge is assumed about the accuracy of these identifications.

\begin{remark}\label{R1}
Note that our approach is independent of how one defines the term \textit{coverage}. For instance, it can be defined in terms of a threshold on the signal-to-interference-plus-noise ratio (SINR) averaged over a certain amount of time in a certain band or averaged over a collection of different bands. However, this is not the focus of our paper. In other words, we model the cell association problem as a nonlinear boundary detection problem and take a machine-learning based approach to solve this problem. This approach is \textit{robust} with respect to erroneous declaration by individual sensor nodes and the irregular shape of cell boundaries regardless of the specific criteria for these declarations. 
\end{remark}

Let us now denote by $y_i$ the declaration of sensor $i$, assumed to have three possibilities $y_i\in\{-1,0,1\}$. In practice, the boundaries are unknown and the objective is to find boundary candidates resulting in minimum detection errors. In machine learning (ML), this problem is equivalent to solving a $3$-class classification such that the following mis-classification error 
$$
\ex{\sum_{i=1}^n \ind{y_i\neq \hat{y}_i}},
$$ 
is minimized, where $\ind{\cdot}$ is the indicator function and $\hat{y}_i$ is the predicted coverage by the classifier based on the training data provided by sensors, i.e., the set $\{(\xb_{i},y_{i})\}_{i=1}^{n}$. The above objective function is non-smooth, and the problem is re-formulated as minimizing a risk functional $R(f)$, defined as \cite{friedman2001elements}
\begin{align}\label{eq:risk}
R(f)&\triangleq\E_{p(\xb,y)}\left[L(f(\xb),y)\right] ~~~~
\Rh(f)\triangleq\frac{1}{n}\sum\limits_{i=1}^{n} L(f(\xb_{i}),y_{i}),
\end{align}
where $L$ is a specific loss function (e.g., $L(y,y')=\max\{0,1-yy'\}$ for Support Vector Machine (SVM) when $y$ is binary), and the expectation is taken with respect to the data distribution $p(\xb,y)$. As $p(\xb,y)$ is unknown, we can only minimize the empirical risk $\Rh(f)$, instead of the true risk $R(f)$, and calculate the gap between the two using standard arguments from measures of function space complexity (e.g., Vapnik–Chervonenkis (VC) dimension \cite{friedman2001elements}, Rademacher complexity \cite{bartlett2002rademacher}, etc). To minimize the risk functional, we need to assume a function class for $f$. Although current ML literature has focused on {\it deep} learning methods for such modeling, they often involve a huge number of parameters. This is an unnecessary complication for the boundary detection problem which is already in a low-dimensional manifold (in this case sensors are located in $\R^2$ plane). Another popular approach in ML and Statistics is {\it kernel} method \cite{hofmann2008kernel}, where 
\begin{align}\label{eq:kernelclass}
    f(\xb)\approx\sum\limits_{i=1}^n \alpha_i k(\xb_i,\xb),
\end{align}
and $k$ is a symmetric positive-definite function called a kernel. The coefficients $\{\alpha_i\}_{i=1}^n$ are unknown and will be learned via empirical risk minimization. Kernel-based boundary detection was proposed by \cite{ding2013kernel} for the case of one BS. However, {\it off-the-shelf} kernel methods are not suitable to solve the problem since they scale prohibitively with respect to the number of sensors $n$. In particular, minimizing the empirical risk $\Rh(f)$ (i.e., optimizing over $\{\alpha_i\}_{i=1}^n$) with kernel methods requires $O(n^2)$ in space and $O(n^3)$ in time \cite{friedman2001elements}. Also, the choice of kernel to use for modeling is a {\it key} decision, which naturally depends on data. In this paper, we are interested in the following problem:

\begin{problem}\label{P1}
Propose a {\it data-driven} solution for kernel selection that improves the boundary detection, i.e., one that dominates {\it data-independent} kernel methods for the task of boundary detection. 
\end{problem}

\section{Data-Driven Random Features}

Despite the popularity of kernel methods for approximation, their poor scalability with respect to the size of data has limited their application in large-scale learning. To improve the {\it computational efficiency}, we use randomized approximation \cite{rahimi2007random}, focusing on kernels of the form  
\begin{align*}
    k(\xb,\xb')&=\int_{\Omega} \phi(\xb ,\omegab)\phi(\xb',\omegab)d\tau(\omegab)\\ &\approx \frac{1}{M}\sum\limits_{m=1}^{M}\phi(\xb ,\omegab_{m})\phi(\xb',\omegab_{m}),\numberthis\label{eq:kernel}
\end{align*}
where $\phi$ is an {\it activation} function (also called basis), and $\{\omegab_{m}\}_{m=1}^{M}$ are independent samples (called {\it random features}) from a given distribution $\tau(\omegab)$ (Monte-Carlo sampling). 

A wide variety of kernels can be approximated via \eqref{eq:kernel} (see e.g., \cite{yang2014random}). Table \ref{tab:distribution table} presents a number of common kernel functions $k(\xb,\xb')$ and their corresponding sampling distributions $\tau(\omegab)$. Observe that these kernels are generally defined for $\xb\in\R^d$, but in this paper $d=2$ as the sensors are located in a 2D area. $\omegab(l)$ (respectively, $\xb(l)$) denotes the $l$-th element of the vector $\omegab$ (respectively, $\xb$). Unbiased kernel estimators are formed with random features sampled from these distributions and evaluated on a cosine feature map, except for the linear kernel where $\phi(\xb,\omegab)=\inn{\xb,\omegab}$. 

\begin{table}
    \centering
    \caption{Different kernel functions and the corresponding sampling distributions of random features.}
    \resizebox{0.4\textwidth}{!}{
    \begin{tabular}{|c||c|c|}
    \hline\xrowht[()]{10pt}
         {\bf Kernel}& $k(\xb,\xb')$& $\tau(\omegab)$\\ 
         \hline\xrowht[()]{20pt}
         Gaussian&
         $e^{-\frac{\norm{\xb-\xb'}_2^2}{2}}$& $(2\pi)^{-\frac{d}{2}}e^{-\frac{\norm{\omegab}_2^2}{2}}$\\ 
         \hline\xrowht[()]{20pt}
         Linear&
         $\inn{\xb,\xb'}$& $(2\pi)^{-\frac{d}{2}}e^{-\frac{\norm{\omegab}_2^2}{2}}$\\ 
         \hline\xrowht[()]{20pt}
         Laplacian & $e^{-\norm{\xb-\xb'}_1}$& $\Pi_{l=1}^d\frac{1}{\pi(1+\omegab(l)^2)}$\\
         \hline\xrowht[()]{20pt}
         Cauchy& $\Pi_{l=1}^d\frac{2}{1+(\xb(l)-\xb'(l))^2}$& $e^{-\norm{\omegab}_1}$\\ 
         \hline
    \end{tabular}}
    \label{tab:distribution table}
\end{table}

Since the main focus of this paper is on the Gaussian kernel, note that we can use the following approximation 
\begin{align*}
k(\xb,\xb') &= e^{-\frac{\sigma^2\norm{\xb-\xb'}_2^2}{2}}\\
&\approx \frac{1}{M}\sum\limits_{m=1}^M 2\cos(\nub_m^\top\xb+b_m)\cos(\nub_m^\top\xb'+b_m),
\end{align*}
where $\{\nub_m\}_{m=1}^M$ come from a multi-variate Gaussian distribution $\Nc(\0,\sigma^2\Ib_d)$ and $\{b_m\}_{m=1}^M$ come from a uniform distribution $\Uc(0,2\pi)$. In this case, the function class will take the form
\begin{equation}\label{RF}
    f(\xb) \approx \sum_{m=1}^M \theta_m\sqrt{2}\cos(\nub_m^\top\xb+b_m),
\end{equation}
where the unknown parameters $\thetab=[\theta_1,\ldots,\theta_M]^\top$ will be learned by minimizing the empirical risk \eqref{eq:risk}. The above approximation is also called a {\it shallow} network \cite{rahimi2009weighted}, provably far more efficient to train compared to \eqref{eq:kernelclass} \cite{rudi2016generalization}. More specifically, the training would now require $O(nM^2)$ in time, which is linear with respect to sensors and significantly smaller than $O(n^3)$ (kernel methods) when $M\ll n$.

\begin{algorithm}[t]
	\caption{Data-Driven Randomized Features (DDRF)}
	{\bf Input:} 
	sensor data $\{(\xb_{i},y_i)\}_{i=1}^n$, an integer $M_0$, an integer $M<M_0$, variance $\sigma^2>0$.
	\begin{algorithmic}[1]
	\STATE Draw $M_0$ independent samples $\{\nub_m\}_{m=1}^{M_0}$ from a multi-variate Gaussian distribution $\Nc(\0,\sigma^2\Ib_d)$. 
	\STATE Draw $M_0$ independent samples $\{b_m\}_{m=1}^{M_0}$ from a uniform distribution $\Uc(0,2\pi)$.
	\STATE Let $\omegab_m=(\nub_m,b_m)$ and $$\phi(\xb,\omegab_m)=\sqrt{2}\cos(\xb^\top\nub_m+b_m).$$\vspace{-0.4cm}
	\STATE Let $\yb=[y_1,\ldots,y_n]^\top$ and construct the matrix 
    \begin{equation}\label{Q1}
        \Qb=\Zb^\top\yb\yb^\top\Zb,
    \end{equation}
	where $\Zb$ is defined as follows
	\begin{equation}\label{transformed matrix}
	    \Zb \triangleq \frac{1}{\sqrt{M_0}}[\Phib(\omegab_1),\ldots,\Phib(\omegab_{M_0})],
	\end{equation}
	and $\Phib(\omegab)$ is defined as follows
	\begin{equation}
	    \Phib(\omegab) \triangleq [\phi(\xb_1,\omegab),\ldots,\phi(\xb_n,\omegab)]^\top.
	\end{equation}
	\STATE For any $i\in \{1,\ldots,M_0\}$, calculate
	\begin{align}\label{weightq}
	    \qh(\omegab_i)=\frac{[\Qb]_{ii}}{\tr{\Qb}}.
	\end{align}
	Let the new weights be $\qbh=[\qh(\omegab_1),\ldots,\qh(\omegab_{M_0})]^\top$.
	\STATE Sort $\qbh$ and select top $M$ features with highest scores from the pool to construct the transformed matrix $\widehat{\Zb}\in\R^{n\times M}$ according to \eqref{transformed matrix} but with $M$ columns. 
	\end{algorithmic}
	{\bf Output:} 
	The transformed sensor data matrix $\widehat{\Zb}$.
\label{ALGO1}
\end{algorithm}

\subsection{Risk Minimization with 3 Classes}\label{sec:3class}
The logistic loss has the form $L(f(\xb),y)=\frac{1}{1+\exp(-yf(\xb))}$ in \eqref{eq:risk}, designed for binary classification. We follow the one-versus-all principle for the multi-class classification problem at hand. Since we have three classes $\{-1,0,1\}$, we decompose the problem into three binary classification problems (though two binary classifiers would be enough). Each binary classifier detects one class against the other two. Returning to Fig. \ref{transmitters} as an example, if a binary classifier declares that a point is not red, and another binary classifier declares that the point is not green, we can easily assign that point to the blue category. 

\vspace{0.2cm}
\noindent
{\bf Data-driven random features:} We will build on model \eqref{RF} for each binary classifier. A major impediment in exploiting the model is the fact that it implicitly transforms the inputs (location of sensors) to a space with a higher dimension, which results in detection algorithms that are not computationally viable. In other words, the dimension of the new space ($\R^M$) is much larger than the original space ($\R^2$). The underlying mathematical intuition is that Monte Carlo sampling of $\omegab_{m}$ (random features) is purely random, and as a result one only approximates well when $M$ is large, translating into risk minimization over a high-dimensional space. We will use {\it data-driven randomization} to improve the modeling quality.

In particular, we employ a general data-driven score for sampling random features introduced in \cite{wang2019general}, where the initial distribution $\tau(\omegab)$ is re-weighted according to the score function $$s(\omegab)=\sum_{i=1}^n\sum_{i=1}^n\phi(\xb_i,\omegab)[\Bb]_{ij}\phi(\xb_j,\omegab),$$
for each $\omegab\in \Omega$, where the matrix $\Bb$ is a task specific positive semi-definite matrix. Here, we will focus on the case where $\Bb = \yb\yb^\top$, which will recover the energy-based exploration of random features (EERF) \cite{shahrampour2018data}. The method is outlined in Algorithm \ref{ALGO1}. By running this algorithm, we transform (hypothetically) each sensor location from $\R^2$ to $\R^M$ using the matrix $\widehat{\Zb}$, which is the algorithm output. The transformed data will then be fed into a logistic regression model as discussed in Section \ref{sec:3class}.

The modification is \emph{data-driven} in the sense that it explicitly takes into account sensors data $\Dc_n=\{(\xb_{i},y_i)\}_{i=1}^n$ in {\it the sampling stage} as well as in the detection stage. This is in contrast to classical methods (e.g., \cite{rahimi2009weighted}) that are data-independent in the sampling phase.

\section{Numerical Experiments}\label{sec:simulation}
We create an artificial dataset according to Fig. \ref{transmitters}. We randomly distributed $n=2000$ sensors in a 2D grid for training and considered two ground-truth boundaries as presented in the figure. As we can see, the training data (provided by sensors) include a number of false declarations according to the ground-truth coverage boundaries; therefore, the prediction (using new data) cannot be \%100 accurate, i.e., we can only approximate the true boundaries up to some error. 
%Using a multi-class logistic regression with transformed feature vector $\phi(\cdot,\omegab)$ we compared two cases of sampling for $\omegab$: data-independent \cite{rahimi2009weighted} vs. data-dependent (proposed score function). As we can observe in Fig. \ref{comparison}, the data-dependent method dominates the data-independent counterpart in terms of the detection accuracy on the test sensor data ($n_{\text{test}}=1000$ sensor data). We can also see that in the sparse regime (say, $M=4$), there is a significant boost in accuracy ($\sim60\%$ to $\sim80\%$).

\vspace{0.2cm}
\noindent
{\bf Benchmark Algorithms:}
Using a multi-class logistic regression with transformed feature vector $\phi(\cdot,\omegab)$, we compared two scenarios for sampling $\omegab$: data-independent vs. data-dependent. The data-independent methods include random kitchen sinks (RKS) \cite{rahimi2009weighted} and orthogonal random features (ORF) \cite{felix2016orthogonal}, which has been proved to be superior that plain random features. The data-dependent case is simply DDRF in Algorithm \ref{ALGO1}.

\vspace{0.2cm}
\noindent
{\bf 1) RKS \cite{rahimi2009weighted}} with $\{\nub_m\}_{m=1}^M$ sampled from Gaussian distribution $\Nc (0,\sigma^2\Ib)$ and $\{b_m\}_{m=1}^M$ sampled from uniform distribution $\Uc (0,2\pi)$ approximates a Gaussian kernel with kernel width $\frac{1}{\sigma}$. We use this set of random features to transform $\Xb \in \R^{n\times 2}$ to $\Zb \in \R^{n \times M}$. The feature map we use for RKS is $\phi(\xb,\omegab) = \sqrt{2}\cos(\nub^\top\xb + b)$.

\vspace{0.1cm}
\noindent
{\bf 2) ORF \cite{felix2016orthogonal}} with $\{\nub_m\}_{m=1}^M$ sampled from Gaussian distribution $\Nc (0,\sigma^2\Ib)$ and modified through QR decomposition approximates a Gaussian kernel with kernel width $\frac{1}{\sigma}$. We use this set of orthogonal random features to transform $\Xb \in \R^{n\times 2}$ to $\Zb \in \R^{n \times 2M}$. The dimension difference of orthogonal random features is resulted from the two dimensional feature map we use, which is $\phi(\xb,\nub) = [\cos(\nub^\top\xb),\sin(\nub^\top\xb)]$.

\vspace{0.2cm}
\noindent
{\bf Practical considerations:} The variance of random features $\sigma$ is set to be the inverse of mean-distance of $50$-th nearest neighbour (in Euclidean distance) following \cite{felix2016orthogonal}. According to this rule, in this simulation $\sigma=1.58$.
The pool size for data-dependent sampling is $M_0 = 10M$ when $M$ random features are used in the classification. 

\begin{figure}[t]
\centering
\includegraphics[width=0.45\textwidth]{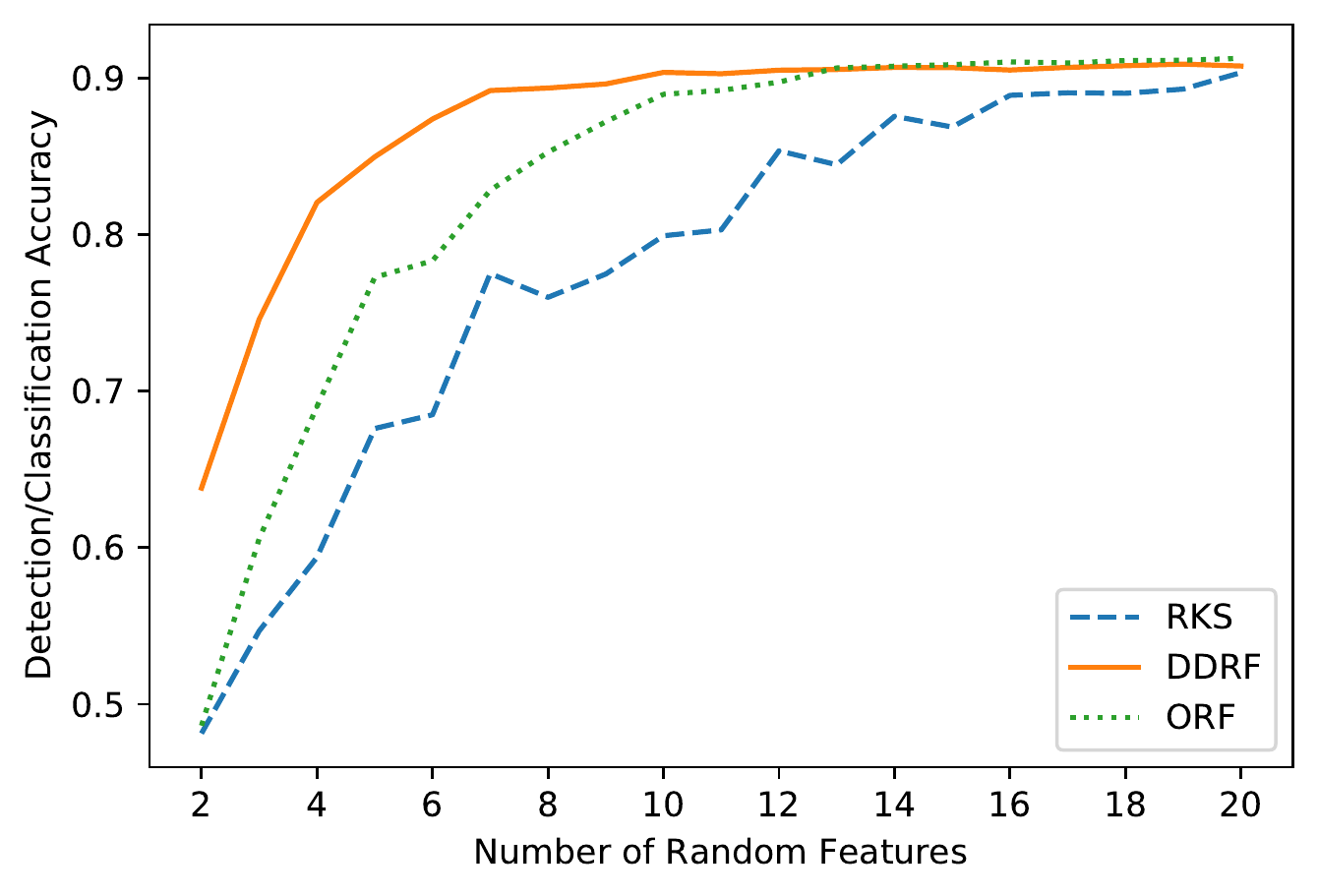}
\caption{DDRF dominates the data-independent counterparts (RKS and ORF) in terms of the detection/classification accuracy.}
\label{comparison}
\end{figure}

\vspace{0.2cm}
\noindent
{\bf Performance:}
The detection accuracy in Fig. \ref{comparison} and the standard error in Table \ref{tab:se} are averaged over 30 simulations. As we can observe in Fig. \ref{comparison}, the data-dependent method dominates the plain random features in terms of the detection accuracy on the test sensor data ($n_{\text{test}}=1000$ sensor data). Although ORF can perform on par with DDRF in the saturated regime, DDRF still shows a significant boost in accuracy in the sparse regime. For example, when $M=4$, the accuracy of DDRF is signifcantly better than RKS ($\sim82\%$ compared to $\sim60\%$ according to Table \ref{tab:se}).

\begin{table}[t]
    \centering
    \caption{The accuracy of DDRF versus two other benchmarks RKS and ORF for various number of random features. The standard errors are reported in parentheses.}
    \resizebox{1\columnwidth}{!}{
    \begin{tabular}{|c||c|c|c|}
         \hline\xrowht[()]{10pt}
         Algorithm &{\tt DDRF}&{\tt RKS}&{\tt ORF}  \\
         \hline\xrowht[()]{8pt}
         $M=4$& {\bf 0.82} (0.011)&0.60 (0.027)&0.68 (0.019)\\
         \hline\xrowht[()]{8pt}
         $M=8$& {\bf 0.90} (0.002)&0.71 (0.027)&0.85 (0.010)\\
         \hline\xrowht[()]{8pt}
         $M=12$&0.90 (0.0008) &0.85 (0.011) &0.90 (0.002)\\
         \hline\xrowht[()]{8pt}
         $M=16$&0.91 (0.0006) &0.88 (0.010) &0.91 (0.0005)\\
         \hline
    \end{tabular}}
    \label{tab:se}
\end{table}

\vspace{0.2cm}
\noindent
{\bf Time cost:} The training time of the three random-feature based algorithms are tabulated in Table \ref{tab: time}. The table also includes the time cost of kernel logistic regression. The training time of random-feature based methods is substantially lower than kernel logistic regression. In particular, the training time of DDRF is roughly \%15.6 of the kernel algorithm. Note that this simulation is only on $n=2000$ sensors, and the difference will be much more significant for larger values of $n$.

\begin{table}[t]
\centering
\caption{The time cost (in seconds) of the algorithms using $M=20$ random features compared with kernel logistic regression.}
\resizebox{0.85\columnwidth}{!}{
    \begin{tabular}{|c||c|c|c|c|}
    \hline\xrowht[()]{10pt}
         Algorithm & {\tt DDRF} & {\tt RKS}  & {\tt ORF} & {Kernel} \\
         \hline\xrowht[()]{8pt}
          Time Cost & 0.036&0.032&0.043& 0.23 \\
    \hline     
    \end{tabular}}
    \label{tab: time}
\end{table}

\section{Conclusion}
 
In this paper, we considered the problem of cell association for cellular users. It is shown how this problem can be modeled as nonlinear boundary detection problem. We then proposed a scalable solution by using randomized shallow networks which utilize data that can be collected by a large number of low-cost spectrum sensors deployed in the field. We also showed how to exploit the power of data-driven modeling in order to reduce the computational cost of training in the proposed solution.

The solution to the boundary detection problem, discussed in this paper, essentially splits the users into two categories, namely, cell-edge users and cell-center users. Eventually, the cell-edge users need to be associated to one of the BSs. The association of cell-edge users to the BSs can be dynamic (real-time). In other words, depending on the network traffic and more importantly the density of nearby users assigned to each of the BSs, the cell association can be different. Developing {\it real-time} classification algorithms to assign cell-edge users to the BSs dynamically is an interesting direction for future work. Furthermore, from a practical point of view, providing a test-bed consisting of radio transmitters and spectrum sensors deployed in the field in order to collect data for training and testing the proposed ML algorithms is another direction for future work.

\addtolength{\textheight}{-0cm}

%%%%%%%%%%%%%%%%%%%%%%%%%%%%%%%%%%%%%%%%%%%%%%%%%%%%%%%%%%%%%%%%%%%%%%%%%%%%%%%%

%\section*{Acknowledgments}
%We gratefully acknowledge the support of Texas A\&M Triads for Transformation (T3) program. 

\bibliographystyle{IEEEtran}
\bibliography{Random-Features,Shahin,hessam}

%%%%%%%%%%%%%%%%%%%%%%%%%%%%%%%%%%%%%%%%%%%%%%%%%%%%%%%%%%%%%%%%%%%%%%%%%%%%%%%%

\end{document}